\documentclass[reprint,amsmath,namssymb,superscriptaddress,longbibliography,aps,aip,prl]{revtex4-1}
\usepackage{graphicx}
\usepackage{dcolumn}
\usepackage{subfigure}
\usepackage{color}
\usepackage{soul}
\usepackage{float}
\usepackage{times}
\usepackage[pdftex,breaklinks,colorlinks,
linkcolor=blue,
anchorcolor=blue,
citecolor=blue,
urlcolor=blue,]{hyperref}

\usepackage{color}
\newcommand{\mpar}[1]{}

\newcommand{\new}[1]{{\color{black} #1}}

\def\be{\begin{equation}}
\def\ee{\end{equation}}
\def\bea{\begin{eqnarray}}
\def\eea{\end{eqnarray}}

\def\r{{\bf r}}
\def\k{{\bf k}}

\def\d{\mbox{d}}

\begin{document}

%\preprint{APS/123-QED}

\title{Observation of structural universality in disordered systems using bulk diffusion measurement}% Force line breaks with \\

\author{Antonios Papaioannou}
\affiliation{City University of New York, The Graduate Center, Department of Physics, New York, NY 10016, USA}

\author{Dmitry S. Novikov}
\affiliation{Center for Biomedical Imaging, Department of Radiology, New York University School of Medicine, New York, NY 10016, USA}

\author{Els Fieremans}
\affiliation{Center for Biomedical Imaging, Department of Radiology, New York University School of Medicine, New York, NY 10016, USA}
 
\author{Gregory S. Boutis}
\email{gboutis@brooklyn.cuny.edu}
\affiliation{City University of New York, The Graduate Center, Department of Physics, New York, NY 10016, USA} 
\affiliation{City University of New York, Brooklyn College, Department of Physics, Brooklyn, NY 11210, USA} 
\date{\today}% It is always \today, today,

\begin{abstract}

%Identifying relevant parameters is central to understanding complex phenomena. This often evokes the concept of universality, which groups microscopically distinct systems into a handful of universality classes, and identifies relevant degrees of freedom effecting their thermodynamic and dynamical properties. 
We report on an experimental observation of classical diffusion distinguishing between structural universality classes of disordered systems in one dimension. Samples of hyperuniform and short-range disorder were designed, characterized by the statistics of the placement of $\mu$m-thin parallel permeable barriers, and the time-dependent diffusion coefficient was measured by NMR methods over three orders of magnitude in time. The relation between the structural exponent, characterizing disorder universality class, and the dynamical exponent of the diffusion coefficient is experimentally verified.  The experimentally established relation between structure and transport exemplifies the hierarchical nature of structural complexity --- dynamics are mainly determined by the universality class, whereas microscopic parameters affect the non-universal coefficients. These results open the way for non-invasive characterization of structural correlations in porous media, complex materials, and biological tissues via a bulk diffusion measurement.

\end{abstract}
\maketitle

How does a measurement of a macroscopic characteristic relate to microscopic structure? This  ill-posed question has been repeatedly asked in many disciplines --- famously, ``Can one hear the shape of a drum?"  \cite{ShapeoftheDrum} --- and its answer depends on the kind of measurement. 
Naively, one could imagine that infinitely many parameters needed to specify a sample's structure would in one way or the other contribute to the  outcome. Physical intuition, however, tells us that only a few parameters  profoundly affect the measurement; identifying these relevant parameters is generally nontrivial, especially for irregular, or {\it disordered} systems. 
For instance, even small irregularities in a periodic lattice can change perfectly conducting metallic bands into an insulator due to quantum localization \cite{gang4}. 

%For instance, even at the resolution level of Fig.~\ref{Fig:Samples}, 
%Regular samples, such as periodic geometries or identical pores, are perhaps the easiest to parametrize. The conceptual difficulty has been presented by {\it structurally disordered} media --- such as a macroscopic imaging voxel of the brain
%Localization -- dimensionality; SR disorder.   

%%%%%%%%%%%%%%%%%%%%%%%%%%%%%%%%%%%%%%%%%%%%%%%%%%%%%%%%%%%
%%%%%%%%%%%%%%%%%%%%%%%%%%%%%%%%%%%%%%%%%%%%%%%%%%%%%%%%%%%
%%%%%%%%%%%%%%%%%%%%%%%%%%%%%%%%%%%%%%%%%%%%%%%%%%%%%%%%%%%

\begin{figure*}[t]
\vspace{-5mm}
\centering
	\includegraphics[width=1.0\textwidth]{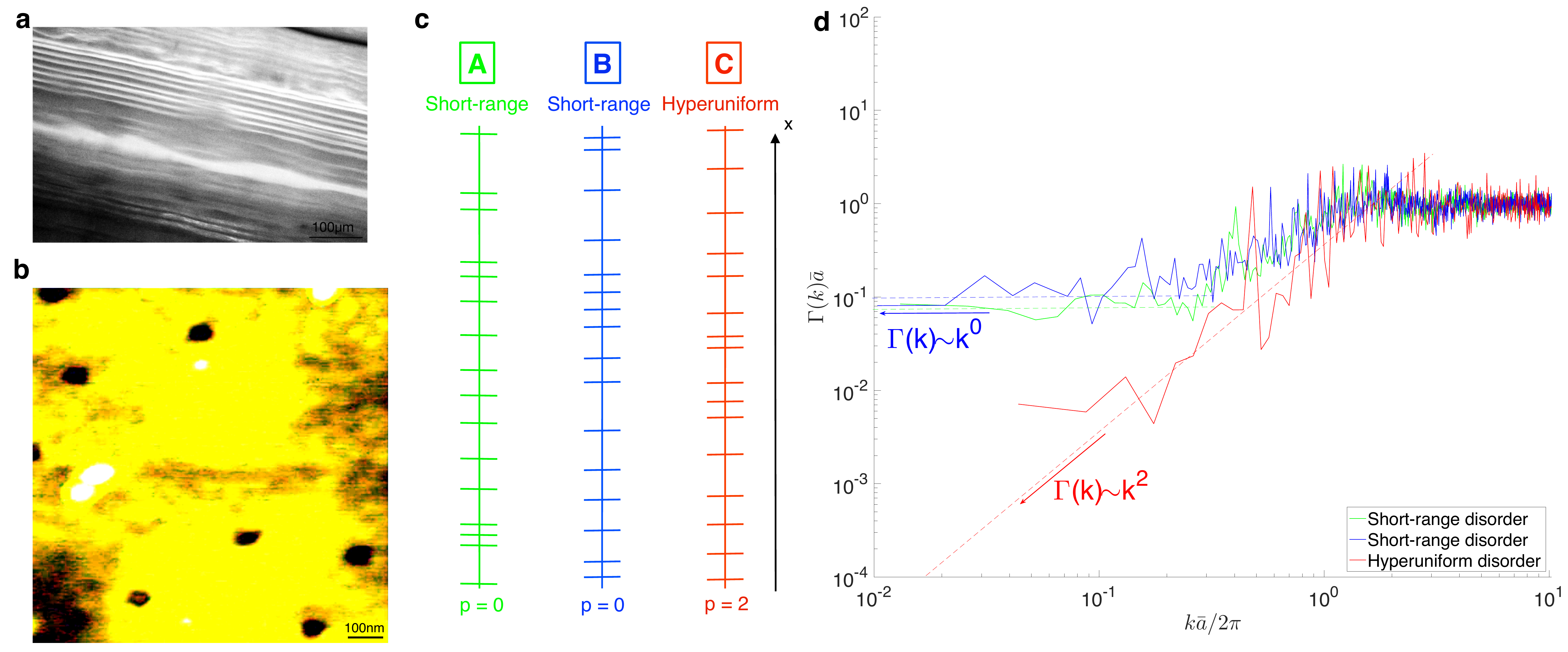}
\vspace{-3mm}
	\caption{\textbf{Structure and universality classes of the samples.} 
	{\bf a} Representative optical microscopy image of the SR sample. 
	{\bf b}, AFM image of a single barrier.
	{\bf c}, Digitized 1d cut-outs of the two SR samples (A-B) and the HU sample C. 
	{\bf d}, Power spectrum (\ref{eq:Correlator}) of the barrier density $n(x)$ reveals qualitative differences between the disorder classes as $k\rightarrow0$:  
	A plateau ($p=0$) in $\Gamma(k)$ for the SR samples (A-B), and $k^p$ scaling with $p=2$ for the HU sample (C).}
	\label{Fig:Samples}
\end{figure*}
 
%%%%%%%%%%%%%%%%%%%%%%%%%%%%%%%%%%%%%%%%%%%%%%%%%%%%%%%%%%
%%%%%%%%%%%%%%%%%%%%%%%%%%%%%%%%%%%%%%%%%%%%%%%%%%%%%%%%%%
%%%%%%%%%%%%%%%%%%%%%%%%%%%%%%%%%%%%%%%%%%%%%%%%%%%%%%%%%%
%%%%%%%%%%%%%%%%%%%%%%%%%%%%%%%%%%%%%%%%%%%%%%%%%%%%%%%%%%
 
Here we consider classical  diffusion in structurally disordered systems, %(Fig. \ref{Fig:Samples}), 
where the practical answer to the above question could help quantify the underlying microstructure of complex materials  \cite{callaghan,PhysRevLett.83.3324,SongMultiple,torquato2013random,karger2014microimaging,Monson2006sorption} and living tissues  \cite{novikov2011random,novikov2014revealing,Burcaw201518,fieremans2016,jones-book}. 
%with diffusion NMR or MRI orders of magnitude below the nominal imaging resolution, with a potential to develop markers of structural integrity and pathology. \cite{novikov2011random,novikov2014revealing,Burcaw201518,fieremans2016} 
We experimentally demonstrate that the qualitative behavior of the time-dependent diffusion coefficient is tied to the long-range structural fluctuations. 
While systems may strongly differ in their microscopic parameters, there are only a few {\it universality classes} of such fluctuations --- in essence, a system can be disordered in one of a few distinct ways --- and each universality class yields a particular power-law behavior of the observed macroscopic diffusion coefficient. % \cite{novikov2014revealing}  

\begin{figure*}[t]
\centering
	\includegraphics[width=1.0\textwidth]{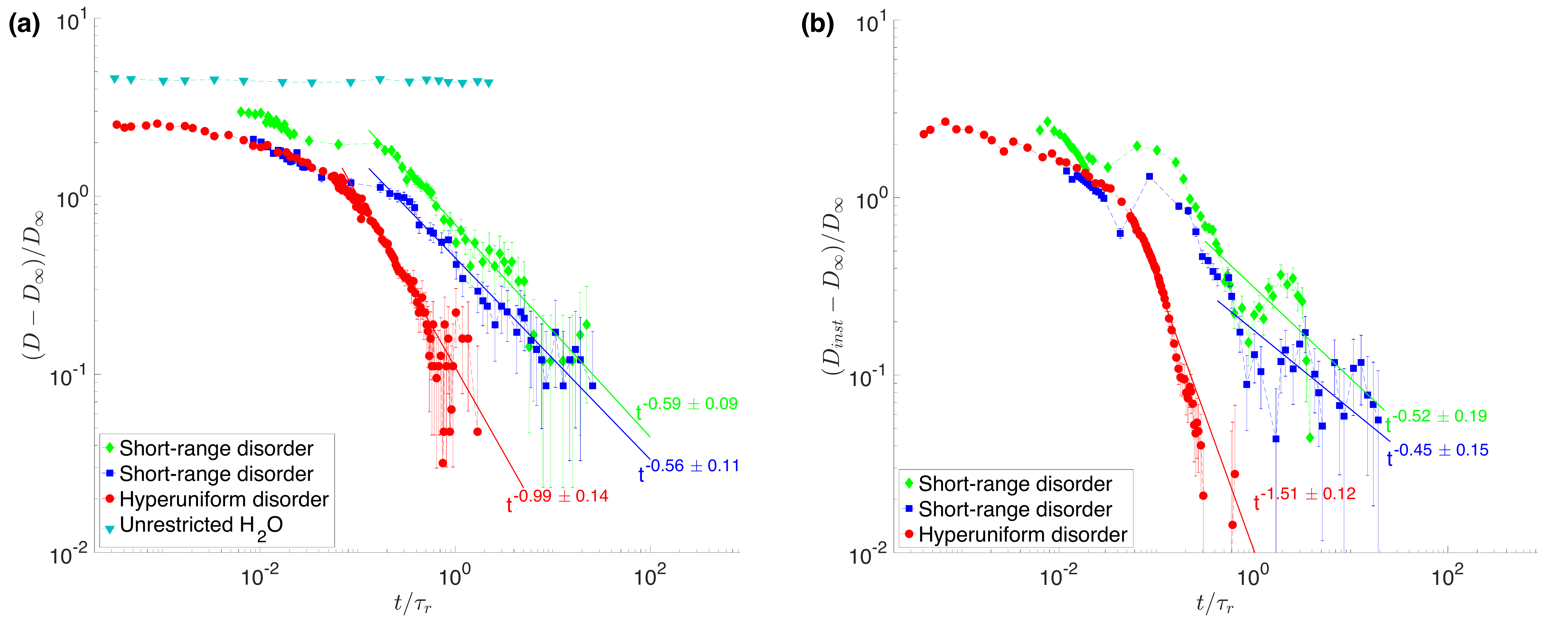}
	\caption{\textbf{Dynamical exponent (\ref{dynamicalexp}) identifies the disorder classes.} 
	{\bf a}, The tail in the cumulative diffusion coefficient $D(t)$ (see text) distinguishes between SR and HU disorder, via exponent 
	$\tilde\vartheta = {\rm min\, }\{\vartheta,\ 1\}$ (Table ~\ref{tab}). Note that $\tilde\vartheta \equiv \vartheta \approx 1/2$ for both SR samples (made of barriers with different permeability), while $\tilde\vartheta \approx 1$ for the HU sample, indicating that the ``true" $\vartheta > 1$. 
	$D(t)=\mathrm{const}$ for unrestricted water is shown for comparison. 
	{\bf b}, To access $\vartheta$ for HU disorder, we obtain the tail in $D_{\rm inst}(t)$, equation (\ref{eq:Dinst}). While results are noisier due to numerical differentiation, the exponent values $\vartheta \approx 1/2$ for SR and $\vartheta \approx 3/2$ for HU, cf. Table~\ref{tab}, are consistent with equation (\ref{dynamicalexp}).
	}
	\label{Fig:Main}
\end{figure*}

Technically, we experimentally verify the recently derived relation  \cite{novikov2014revealing}
\begin{equation}
\label{dynamicalexp}
\vartheta=\frac{p+d}{2}
\end{equation}  
between the structural exponent $p$, and the dynamical exponent $\vartheta$ of the Brownian motion $x_t$ 
in structurally disordered stationary media in $d$ spatial dimensions. The defining signature of structural complexity is reflected in the structural exponent $p$ which takes discrete values according to the universality class, as illustrated in Fig.~\ref{Fig:Samples} for our $(d=1)$-dimensional samples. 
Equation (\ref{dynamicalexp}) relates $p$ to the long-time tail in the bulk diffusion coefficient 
  \cite{novikov2014revealing} (the mean-squared displacement rate)
\begin{equation} \label{eq:Dinst}
D_{\rm inst}(t)\equiv\frac{\partial}{\partial t}\frac{ \langle (x_t-x_0)^2 \rangle}{2}
\simeq D_{\infty}  \, + \,  c\cdot t^{-\vartheta}\,, \quad t \to \infty .  
\end{equation}
The macroscopic diffusion coefficient $D_{\infty} \equiv D_{\rm inst}(t)|_{t=\infty}$ and the power-law amplitude $c$ are non-universal, i.e. depend on the microstructural parameters. On the other hand, 
as we experimentally demonstrate in Fig.~\ref{Fig:Main}, 
the relation (\ref{dynamicalexp}) is universal \cite{novikov2014revealing}, akin to the relations between critical exponents  \cite{RevModPhys.49.435} in statistical physics,  where the notion of universality originates.

Formally, the structural universality class 
is defined \cite{novikov2014revealing} by the $k\to0$ scaling of the power spectrum
\begin{equation}
\label{eq:Correlator}
\Gamma(k)\equiv \int_V\! \d\r\, e^{-i\k\r} \langle n(\r_0+\r)n(\r_0) \rangle_{\r_0}=\frac{\lvert n(\k)\rvert^2}{V} \sim k^p 
\end{equation}
of the restrictions which embody the sample's microscopic structure. 
The exponent $p$, taking a handful of discrete values such as in Fig.~\ref{Fig:Samples}d, describes how fast the spatial correlations $\Gamma(\r)$ in the density of the restrictions $n(\r)$ decay  at large distances $r$, and thereby characterizes the system's heterogeneity.
The values $p>0$ correspond to {\it hyperuniform} media \cite{PhysRevE.68.041113,1742-5468-2009-12-P12015} (sample C), where the fluctuations are suppressed relative to the short-range (e.g. Poissonian) disorder ($p=0$, samples A and B); $p<0$ correspond to {\it strong disorder}, where the fluctuations are enhanced \cite{novikov2014revealing,novikov2011random}. Qualitatively, the variance in the number of restrictions within a volume $V$ grows $\propto V$ for short-range disorder (according to the central limit theorem), slower than $V$ for hyperuniform disorder (such as in maximally random jammed packings \cite{zachary2011}), and faster than $V$ for strong disorder. 
The relation (\ref{dynamicalexp}) relies on self-averaging  \cite{PhysRevLett.77.3700}, $p+d>0$, ensuring the existence \cite{novikov2014revealing} of finite $D_{\infty}$.

Two samples exhibiting short-range (SR) disorder were constructed by stacking flat, porous, permeable barriers in a layered geometry (as shown in Supplementary Fig. S1c \cite{Supplemental}), and random positions, inside a glass tube filled with H$_2$O. One SR sample was constructed using the barriers with 15 $\rm nm$ pore diameter (A in Fig. \ref{Fig:Samples}c) and one SR sample using the barriers with 45 $\rm nm$ pore diameter (B in Fig. \ref{Fig:Samples}c). Fig. \ref{Fig:Samples}b reveals a pore density of 8 pores/$\mu m^2$ by AFM. These two different samples correspond to two different realizations of short-range disorder and the one-dimensional lines shown in Fig. \ref{Fig:Samples}c correspond to digitized cut-outs of the actual samples \mpar{c1}\new{representing the barrier spacings of a part of the sample}.

A representative optical microscopy image of SR sample A is shown in Fig.~\ref{Fig:Samples}a and yields an average spacing $\bar{a}\approx 12.5 \,\mu$m between the centers of the barriers. The short-range character of the arrangement is proven by the finite value of the plateau $\Gamma(k)|_{k\to0}$ of the power spectrum, Fig.~\ref{Fig:Samples}d, and is also consistent with the \mpar{a.1}\new{probability density function (PDF)} of the successive barrier spacings (Supplementary Fig.~S7 \cite{Supplemental}) lacking a ``fat tail''. The non-Possonian nature of barrier arrangement in both SR samples is shown by the value $\Gamma(k)|_{k\to0}\cdot \bar a $  which  is different from unity (in contrast to the Poissonian, i.e. fully uncorrelated placement), and is consistent with non-exponential PDF of the barrier spacings.

%\new{For the computation of $\Gamma(k)$, the optical microscopy image was digitized resulting in 1d disordered lines of length $L$. The resulting 1d lines where then concatenated and randomized to create a pseudo-averaging among disorder realizations (see Supplementary section II for detailed description as well as Fig. S8)). }
%This hypothesis is supported by non-Poissonian PDF of the barrier spacing of both SR samples shown in Fig. S7 of the Supplementary. A purely Poissonian placement of the barriers would yield a $\Gamma(k)= 1$ plateau for low-$k$ values along with an exponential PDF, revealing no correlations between the barrier placements. However, the experimentally determined power spectra (eq. \ref{eq:Correlator}) for both samples, shown in Fig. \ref{Fig:Samples}e, exhibit a finite plateau $\Gamma(k)\sim k^0$, lower than unity, as $k\to 0$, corresponding to the structural exponent $p=0$. These obervations would point towards short-range disorder revealing correlations in the barrier placement. In addition the PDF shown in Fig. S7 reveals a low probability of large gap occurrence in the barrier placement which would point to a distribution with a fat tail \cite{novikov2014revealing} and therefore different behavior of $\Gamma(k)$ and dynamics.

%\mpar{dont you want to show custom made coils, indicate challenges of manufacturing all this stuff? Maybe in Methods?}
On the other hand, the hyperuniform (HU) disordered sample C, shown in Fig.~\ref{Fig:Samples}c (and Supplementary Fig. S1-a-b-d \cite{Supplemental}), was achieved by placing identical rectangular copper plates, $\sim(45\pm4)$ $\rm \mu m$ thick, between the permeable barriers with pore diameter of 45 $\rm nm$ and is characterized by reduced long-range structural fluctuations. Ideally, the barriers would create a periodic lattice (with $\bar{a}\approx 51.0 \,\mu$m) which would result in Bragg peaks in $\Gamma(k)$ and $\Gamma(k<\pi/\bar a)\equiv 0$. However, experimental inaccuracies in the placement of the barriers and copper plates act as random displacements from ideal lattice positions, resulting in apparent hyperuniformity \cite{PhysRevE.68.041113} of a ``shuffled lattice" \cite{gabrielli2003}, for which 
the power spectrum $\Gamma(k)\sim k^2$ for $k\bar a\ll 1$.
%(cf. equation [S22] in ref.~ \cite{novikov2014revealing}). 
%Reduced fluctuations are highlighted by the power spectrum, shown in 
The spectrum in Fig.~\ref{Fig:Samples}d is indeed consistent with the exponent value $p=2$.   

%%%%%%%%%%%%%%%%%%%%%%%%%%%%%%%%%%%%%
%%%%%%%%%%%%%%%%%%%%%%%%%%%%%%%%%%%%%
 %%%%%%%%%%%%%%%%%%%%%%%%%%%%%%%%%%%%%
 %%%%%%%%%%%%%%%%%%%%%%%%%%%%%%%%%%%%%
 %%%%%%%%%%%%%%%%%%%%%%%%%%%%%%%%%%%%%

We underscore that it is practically impossible to discern the qualitative differences between the samples A, B and C --- or to reveal the disorder universality class by the naked eye. Based on local sample cut-outs, shown in Fig.~\ref{Fig:Samples}c, the three samples look very similar, when the dimensions are rescaled such that mean spacing between the barrier centers is the same for all of them. 
However the power spectrum $\Gamma(k)$, shown in Fig.~\ref{Fig:Samples}d, readily shows similarity between samples A and B, and their qualitative difference from sample C, as its low $k$ scaling captures the universal features in the large-scale behavior of the density fluctuations.  \mpar{A1-B1-C}\new{For the computation of $\Gamma(k)$, the reader is referred to Supplementary section II as well as Fig. S8. \cite{Supplemental}}
%Note that the power spectra shown in Fig.~\ref{Fig:Samples}e are rescaled with $\bar{a}$ which corresponds to the HU sample (C). 
In what follows, we show how a bulk diffusion measurement distinguishes between the SR and HU classes, thereby yielding the form of $\Gamma(k)$ for $k\bar a \ll 1$ (i.e. for distances exceeding $\bar a$), 
and experimentally validating the relation (\ref{dynamicalexp}) in dimension $d=1$.

%To address the above question, first we experimentally determine $\Gamma(k)$ (eq. \ref{eq:Correlator}) of the three phantoms using the optical microscopy images shown in Fig. \ref{Fig:Samples}a-b. The latter correlator is shown in Fig. \ref{Fig:Samples}e. A plateau in $\Gamma(k)\sim k^0$ is observed as $k\rightarrow0$ for the phantoms in green and blue, corresponding to a structural exponent of $p=0$ verifying the expected short-range disorder. On the other hand, a power law scaling of $\Gamma(k)\sim k^2$ was observed for the phantom in red corresponding to a structural exponent of $p=2$ implying hyperuniformity. Note that the fluctuations at high $k$ values of $\Gamma(k)$ result from the local barrier structure profile in signal processing. The two experimentally determined structural exponents, $p$, set the expected dynamical exponents via eq. \ref{dynamicalexp} based on the theory. For short-range disorder in one dimension $p=0$, and according to eq. \ref{dynamicalexp}, $\vartheta\rvert_{SR}=1/2$. On the other hand for hyperuniform disorder, $p=2$, and $\vartheta\rvert_{HYP}=3/2$.

%Remarkably, the measured diffusion coefficient (\ref{eq:Dinst}) exhibits qualitatively distinct behavior and is able to tell between the universality classes. 
 
%To obtain the instantaneous water diffusion coefficient (\ref{eq:Dinst}), 
The conventional {\it cumulative} $D(t) \equiv \langle (x_t-x_0)^2\rangle/2t$ of H$_2$O was measured using pulse-gradient diffusion NMR \cite{callaghan} over a broad  
%investigate the transport dynamics of H$_2$O through the phantoms the cumulative diffusion coefficient, $D$, was measured which is a metric of mean square molecular displacements and is related to the instantaneous diffusion coefficient via,
%\begin{equation}
%D(t)\equiv\frac{ \langle \delta z^2 \rangle}{2t}=\frac{1}{t}\int_0^tdt'D_{inst}(t').
%\label{Dinversetime}
%\end{equation}
%Both the cumulative and instantaneous diffusion coefficients were measured for a wide 
range of diffusion times $t$, from 1.0\,ms to 4.5\,s, spanning over 3 orders of magnitude, and translating to mean square displacements $ \langle (x_t-x_0)^2\rangle^{1/2}$ ranging from 2\,$\mu$m to 144\,$\mu$m. Measuring such short mean square displacements requires fast switching and strong in magnitude gradient pulses. Therefore, a homemade gradient coil was constructed \cite{suits1989improving,zhang1998pulsed} capable of delivering gradient pulses of approximately 90 $\mathrm{G/cmA}$. However, such strong gradient pulses may introduce errors in the experimental data, such as those due to eddy currents. To mitigate such effects, two pulse sequences were used for the diffusion measurements (cf. Supplementary Materials \cite{Supplemental}) which made use of bipolar gradient pulses for short times, and asymmetric pulses for long times. 
%\mpar{1-2 sentences about challenges of measurements. 
%Each experimental run took HOW MANY DAYS?}
%The diffusion coefficient (\ref{eq:Dinst}) was futher obtained via $D_{\rm inst}(t) = \partial [t D(t)]/\partial t$, using numerical differentiation with Savitzky-Golay regularization \cite{Savitzky}, see Methods section. 
%This wide dynamic range enabled us to probe both the short- and long-time transport dynamics in the three phantoms and directly observe their universality class via the dynamical exponent, $\vartheta$, equation~\ref{eq:Dinst}.  

Figure \ref{Fig:Main}a shows the time dependence of the cumulative diffusion coefficient $D(t)$, of H$_2$O diffusing through the three samples, as well as for  
unrestricted H$_2$O (cyan). Note that the diffusion coefficient for unrestricted H$_2$O (cyan) was rescaled using $D_{\infty}$ from sample A. While there is no time dependence in $D(t)$ for  unrestricted H$_2$O,  a power-law exponent $\tilde\vartheta=0.59\pm0.09 $ in $D(t) - D_\infty \sim t^{-\tilde\vartheta}$ was observed for H$_2$O diffusing through sample A and $\tilde\vartheta=0.56\pm0.11 $ for sample B. Note that the exponent $\tilde\vartheta$ is the same with $\vartheta$ of eq. (\ref{eq:Dinst}) if $\vartheta<1$. The exponents are in remarkable agreement with equation (\ref{dynamicalexp}) for $p=0$ and $d=1$, 
and with earlier prediction \cite{Ernst-I} for the tail in $D(t)$. On the other hand $D(t)-D_\infty$ for H$_2$O diffusing through the HU sample exhibits the $1/t$ tail with $\tilde\vartheta=0.99\pm0.14$.\mpar{a.3} \new{The range in which the least squares fit was performed was chosen such that the $\chi^2/\mathrm{dof}$ was minimized.} The structural and dynamical exponents, as well as the main characteristic of the samples such as residence and diffusion times $\tau_r$ and $\tau_D$, are given in Table~\ref{tab}. 

%%%%%%%%%%%%%%%%%%%%%%%%%%%%%%%%%%%%%
 %%%%%%%%%%%%%%%%%%%%%%%%%%%%%%%%%%%%%
 %%%%%%%%%%%%%%%%%%%%%%%%%%%%%%%%%%%%%
 %%%%%%%%%%%%%%%%%%%%%%%%%%%%%%%%%%%%%
\begin{table*}[t] 
\centering
%\scriptsize
 \begin{tabular}{ l | c | c | c  c |  c  c  |c|c|c || c  c}
 %\hline
 Sample & Disorder class & $p$ & $\tilde\vartheta_{\rm th}$ & $\tilde{\vartheta}$ & \ \ $\vartheta_{\rm th}$ \ \ &\ \ $\vartheta$ \ \ & $\tau_r$, ms&$\tau_D$, ms&$D_{\infty}, \frac{\rm  \mu m^2}{ms}$&$\bar a$, $ \rm \mu$m \ \ &\ \ Pore diam., nm \\ [0.5ex] 
 \hline
 A (green-diamonds) &SR & 0 & $1/2$ & $0.59\pm0.09$& $1/2$ &$0.52\pm0.19$&157.2&34.2&$0.42\pm0.04$& 12.5 & 15 \\ 
 %\hline
 B (blue-squares) &SR & 0 & $1/2$ & $0.56\pm0.11$ &$1/2$  &$0.45\pm0.15$&117.2&43.0&$0.58\pm0.05$& 14.1 &45\\
 %\hline
 C (red-circles) &HU & 2 & $1$ &$0.99\pm0.14$ & $3/2$  &$1.51\pm0.12$&2949.0&754.9&$0.63\pm0.04$& 58.9 &45\\[1ex]
%  \hline
\end{tabular}
\caption{ \label{tab}  {\bf Sample parameters and exponents for disorder classes.} Theoretical (eq. \ref{dynamicalexp}) and experimental (Fig.~\ref{Fig:Main}) power-law exponents $\tilde\vartheta$ and $\vartheta$ in the tails of $D(t)$ and $D_{\rm inst}(t)$. 
The (non-universal) macroscopic diffusion coefficient, $D_{\infty}$, mean barrier spacing $\bar{a}$ \new{as computed from optical microscopy}, pore diameter of the barriers, 
residence time $\tau_r\equiv\bar{a}/2\kappa$, and time to diffuse  in-between barriers, $\tau_D\equiv\bar a^2/2D_0$ are also shown. }%For phantoms A,B,C the reader is referred to Fig.\ref{Fig:Samples}.
\label{Table}
\end{table*}
%%%%%%%%%%%%%%%%%%%%%%%%%%%%%%%%%%%%%
 %%%%%%%%%%%%%%%%%%%%%%%%%%%%%%%%%%%%%
 %%%%%%%%%%%%%%%%%%%%%%%%%%%%%%%%%%%%%
 %%%%%%%%%%%%%%%%%%%%%%%%%%%%%%%%%%%%%
%%%%%%%%%%%%%%%%%%%%%%%%%%%%%%%%%%%%%

The $1/t$ tail in $D(t)$ in the HU sample indicates that $\vartheta>1$. Indeed, the cumulative $D(t) \equiv \frac1t\, \int_0^t\! \d \tau\, D_{\rm inst}(\tau)$ may be used to determine $\vartheta$ only in the case when the power-law tail in $D_{\rm inst}(t)$ is sufficiently slow \cite{novikov2014revealing}, $\vartheta < 1$. In this case, the instantaneous mean squared displacement rate (\ref{eq:Dinst}) has similar behavior to the average rate  $\langle (x_t-x_0)^2\rangle/2t$ over the whole interval $t$; formally, the above integral converges at the upper limit. 
However, when the underlying $\vartheta >  1$, the tail  
$D(t)-D_\infty  = \frac1t\, \int_0^t\! \d\tau\, [D_{\rm inst}(\tau)-D_\infty] \simeq \frac1t\,  \int_0^\infty\! \d\tau\, [D_{\rm inst}(\tau)-D_\infty]$ is determined by the \textit{short} $\tau$, such that the $1/t$ factor overshadows the effect of $\vartheta$. In other words, $D(t)-D_\infty \sim t^{-\tilde\vartheta}$, where $\tilde\vartheta = \mathrm{min\, } \{\vartheta,  \ 1\}$. Hence, if the tail in $D(t)$ has $\tilde\vartheta = 1$, which is the case for the HU sample, one has to obtain $D_{\rm inst}(t)$ via numerical differentiation to uncover the true $\vartheta > 1$, with the expense of amplifying the experimental noise.

%and H$_2$O diffusing through the three constructed phantoms. We observe no time dependence in the cumulative diffusion coefficient, $D$, of unrestricted H$_2$O whereas a characteristic time dependence of
%\begin{equation}
% D(t)\simeq D_{\infty} + t^{-\tilde{\vartheta}},
% \label{Maskedexp}
% \end{equation}
%was observed for H$_2$O diffusing through the phantoms as a result of disorder. Note that the definition \ref{Dinversetime} may mask the true dynamical exponent; if $\vartheta>1$ as it is the case for hyperuniform disorder, a power law exponent $\tilde{\vartheta}=1$ will become apparent in the cumulative diffusion coefficient and one has to compute $D_{inst}$ using definition \ref{eq:derivtD}. On the other hand if $\vartheta<1$, as it is the case for short-range disorder, then $\tilde{\vartheta}=\vartheta$. 

%For short-range disorder (green,blue), and for diffusion times greater than the residence time between the films, $\tau_r\equiv\bar{a}/2\kappa$ ($\tau_{r\mathrm{SR}}\rvert_{15nm} \simeq$ 37.0 $ms$, $\tau_{r\mathrm{SR}}\rvert_{45nm} \simeq$ 23.0 $ms$), we observe a dynamical exponent of $\tilde{\vartheta}=0.59\pm0.09\rvert_{15nm}$ and $\tilde{\vartheta}=0.56\pm0.11\rvert_{45nm}$ in agreement with the theory  \cite{novikov2011random,novikov2014revealing}. 

Figure \ref{Fig:Main}b shows the computed instantaneous $D_{\rm inst}(t) = \partial_t [t D(t)]$, using numerical differentiation with Savitzky-Golay (SG) regularization \cite{Savitzky} (cf. Supplementary Materials \cite{Supplemental}), along with the weighted least squares fit (solid line).The time window in which the fit was performed was chosen such that the $\chi^2/\mathrm{dof}$ was minimum.
%shows the computed $D_{inst}$ using the definition \ref{eq:derivtD}. 
As expected, for both SR samples, $D_{\rm inst}(t)$ reaches its universal limit $D_{\infty}$ according to equation~(\ref{eq:Dinst}) with $\vartheta=0.52\pm0.19$ for sample A and $\vartheta=0.45\pm0.15$ for sample B (cf. Table~\ref{tab}), consistent with the above results for $\tilde\vartheta$ and equation (\ref{dynamicalexp}) with $p=0$ and $d=1$. 
%$\vartheta=0.57\pm0.12\rvert_{15nm}$ and $\vartheta=0.53\pm0.15\rvert_{45nm},$ consistent with the scaling of the cumulative diffusion coefficient ($\tilde{\vartheta}\simeq\vartheta$), and the theory. Tabulated results of the observed dynamical exponents, $\vartheta$ and $\tilde{\vartheta}$, structural exponents, $p$, disorder classes and expected power law exponents based on the theory are shown in Table \ref{Table}.
%For $t>\tau_{r\mathrm{HYP}} \simeq$ 162.0 $ms$
%\mpar{check $\vartheta$ in text and in Table to agree} 
For the HU sample, the dynamical exponent $\vartheta=1.51\pm0.12$, is notably different from that for  SR samples, and in agreement with equation (\ref{dynamicalexp}) for $p=2$ and $d=1$. The least squares fit was stable with respect to the SG filtering window and polynomial order producing reasonable values of $\chi^2/\mathrm{dof}$ (cf. Supplementary Materials for details, Fig. S4 and S5 \cite{Supplemental}). \mpar{b2}\new{Note that the fit is mainly weighted by the first points which have good signal-to-noise ratio.}\mpar{A2} \new{An important observation of Figure \ref{Fig:Main}b is that the molecules in the HU sample gets homogenized by the diffusion process qualitatively faster than in the SR samples A-B, so that the power law tail becomes pronounced already when $t\sim \tau_r$. This is a general consequence of a more efficient coarse-graining in a qualitatively more ordered (hyperuniform) sample. As noted in ref.~\cite{novikov2014revealing}, in the ``extreme" case of a fully periodic sample, diffusion exhibits coherence due to infinitely long spatial correlations, which makes the sample effectively homogenized already when $t\sim \tau_D$.}
%a dynamical exponent of $\tilde{\vartheta}=0.99\pm0.14$ was observed for H$_2$O diffusing through the phantom exhibiting hyperuniform disorder as described by definition \ref{Dinversetime}. Using definition \ref{eq:derivtD}, $D_{inst}$ reveals the true dynamical exponent $\vartheta=1.47\pm0.18$, is in excellent agreement with the theory.

Previous  applications of bulk diffusion  for characterizing microstructure below imaging resolution focussed on the {\it short-time} \cite{PhysRevLett.68.3555} initial decrease $D(t)\simeq D_0(1-\frac{4\sqrt{D_0}}{3d\sqrt{\pi}}\frac{S}{V}\cdot t^{1/2})$ of the cumulative diffusion coefficient, as a result of the increasing fraction $\sim \sqrt{D_0 t}\, S/V$ of random walkers  restricted by walls. 
In this limit, it is the {\it net amount} of the restrictions that is relevant, irrespective of their positions in space --- akin to the net drum surface area derived from the density of high frequency eigenmodes \cite{ShapeoftheDrum}. 
This technique has been used for quantifying the surface-to-volume ratio ($S/V$) of porous media  \cite{PhysRevLett.83.3324} and biological samples such as red blood cell suspensions \cite{latour1994time} and brain tumor cells in mice \cite{reynaud2016}.
%, to determine the surface-to-volume ratio $S/V$ of barriers (membranes). 
%For diffusion times smaller than the typical time across a pore of diameter $\bar{a}$, $\tau_D=\bar{a}^2/2D_0$, only a fraction of the diffusing molecules experience restrictions leading to a universal decay of the diffusion coefficient given by $D(t)\simeq D_0(1-\frac{4\sqrt{D_0}}{3d\sqrt{\pi}}\frac{S}{V}t^{1/2})$ \cite{PhysRevLett.68.3555}. 

\begin{figure}[t]
\centering
	\includegraphics[width=0.45\textwidth]{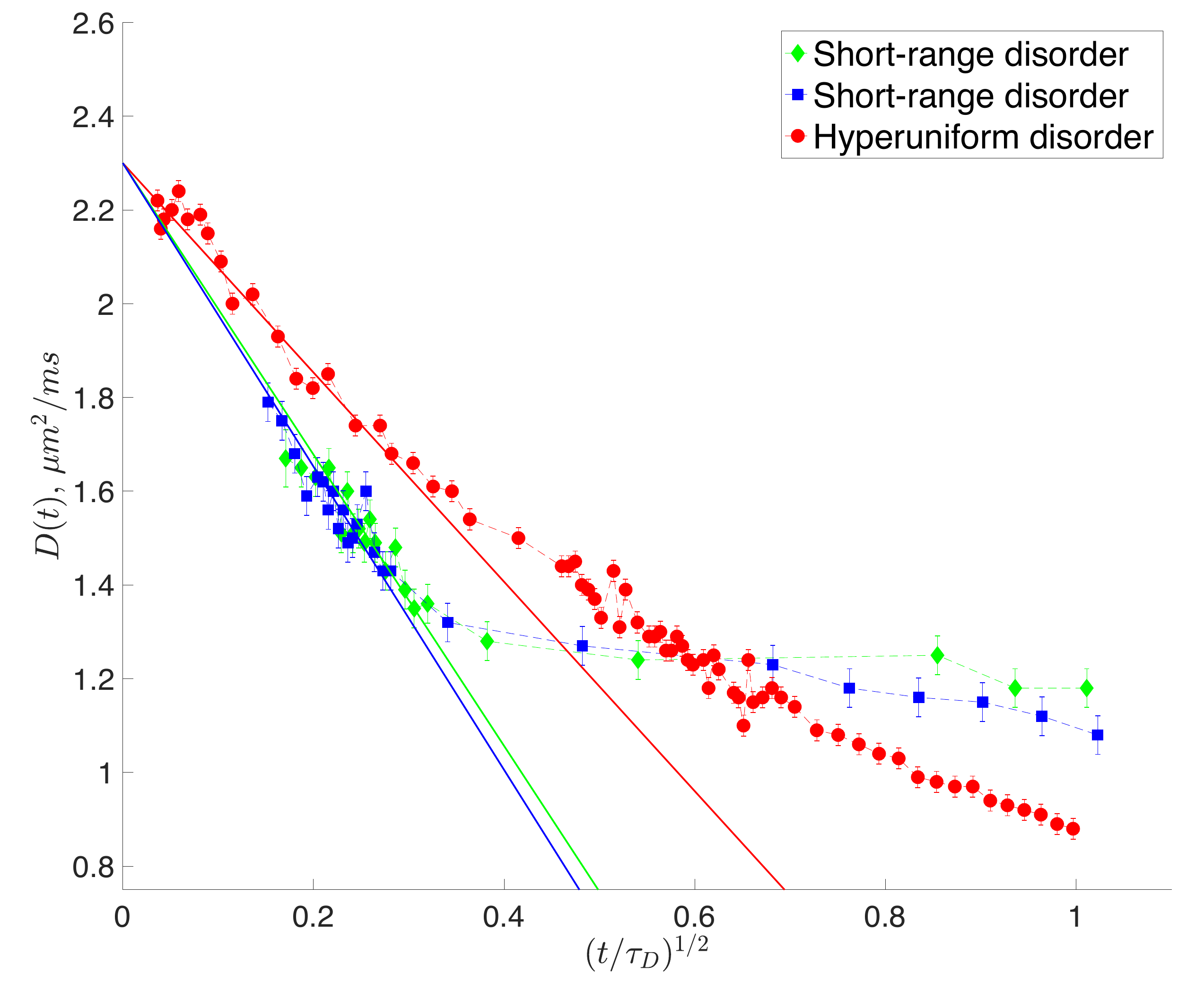}
	\caption{\textbf{Short time $t^{1/2}$ decrease \cite{PhysRevLett.68.3555} of $D(t)$:} 
	Quantifying the net amount of restrictions, $S/V = 2/\bar a$; the difference in the large scale fluctuations is not revealed. 
	 The cumulative diffusion coefficient $D(t)$ exhibits the $t^{1/2}$ decrease for $t/\tau_D\ll1$.   
	}
	\label{Fig:Cumulative_SV}
\end{figure}

Experimentally, the short-time limit is highly demanding on the pulsed field gradients. However, for our samples, displacements as short as $L(t)\approx2$ $\rm \mu m$ are accessible with our homemade gradient coil. Fig.~\ref{Fig:Cumulative_SV} highlights the initial $t^{1/2}$ decrease of $D(t)$ for $t/\tau_D \ll 1$, when the short time limit is valid (cf. Table \ref{Table} for  the values of $\tau_D$). 
%\mpar{ put all in Table: ($\tau_{D\mathrm{SR}}\simeq5.5$ $ms$, $\tau_{D\mathrm{HYP}}\simeq380.0$ $ms$), $S/V$, $\tau_D$, $\tau_r$, etc.} 
For sample A, the average spacing of the barriers was determined from $S/V\equiv2/\bar{a}$, and found to  be $\bar{a}=11.4$ $\mu$m, deviating by $\sim9$\% from the value expected from the images acquired via optical microscopy. Simirarly, for sample \mpar{c3}\new{B}, $\bar{a}=12.0$ $\mu$m deviating by $\sim15$\% from the value expected from the images acquired via optical microscopy and reported in Table \ref{Table}. For HU sample, $\bar{a}=61.5$ $\mu$m  deviated by approximately $\sim4$\% from the predicted value (\new{Table \ref{Table}}). In the least squares fits shown in Fig.~\ref{Fig:Cumulative_SV}, the free diffusion coefficient $D_0$ was fixed to the exact value at the corresponding temperature. Note that the maximum $(t/\tau_D)^{1/2}$ used for the least squares fit (solid lines of Fig.~\ref{Fig:Cumulative_SV}) was 0.31 for sample A, 0.27 for sample B and 0.24 for sample C (see Supplementary Materials \new{Fig. S6} for statistical analysis of the fit \cite{Supplemental}). As mentioned earlier, the initial $t^{1/2}$ decrease, sensitive only to the net amount of restrictions, cannot reveal structural correlations. 
Therefore, the qualitative differences between the two disorder classes are not apparent in Fig.~\ref{Fig:Cumulative_SV} --- note that the difference in the slopes between the curves arises from differences in the $\bar{a}$, entering $\tau_D$, between the optical microscopy and the least squares fit. Here, $\bar{a}$ from optical microscopy was used to compute $\tau_D$.

To summarize, our experiments reveal the qualitative difference in the diffusive dynamics between samples with qualitatively different spatial statistics of structural fluctuations, justifying the application of the concept of universality to classical transport in disordered media, and validating the fundamental relation (\ref{dynamicalexp}) between structural and dynamical exponents. 
%similarities between SR phantoms A and B having the same structural exponent $p$, and dynamical exponent $\vartheta$ verifying the essence of universality. 
The coefficients $c$ and $D_{\infty}$ of equation (\ref{eq:Dinst}) for the two SR samples are non-universal, 
and reflect the density of the barriers and their permeability 
(cf. Supplementary Materials \cite{Supplemental}). %yielding different diffusive permeability values $\kappa$, 
However, the dynamical exponent $\vartheta$ remains the same, because the statistics of large-scale fluctuations for both samples A and B are governed by the central limit theorem (finite correlation length, a plateau in $\Gamma(k)|_{k\to 0}$). On the other hand, based on the dynamical exponent $\vartheta$, qualitative differences were revealed between the samples exhibiting short-range (A, B), and hyperuniform disorder (C) (where fluctuations are reduced \cite{PhysRevE.68.041113,1742-5468-2009-12-P12015} relative to those governed by central limit theorem since $\Gamma(k)|_{k\to 0}\to 0$), verifying that diffusion can identify the structural universality class of the medium.

%%%%%%%%%%%%%%%%%%%%%%%%%%%%%%%%%%%%%
%%%%%%%%%%%%%%%%%%%%%%%%%%%%%%%%%%%%%

After the seminal 1991 observation of diffusion diffraction \cite{Callaghan_N} yielding the structure factor of water-filled identical confining pores,
the late Paul T. Callaghan insightfully referred to diffusion as microscopy \cite{callaghan}. 
This $q$-space technique has enabled determination of the shape of regular confining structures with impermeable walls, such as pores of \mpar{c4}\new{any} shape \cite{laun2011determination}.
The present investigation suggests that 
%a complementary part of the diffusion propagator --- 
the time-dependent diffusion coefficient (\ref{eq:Dinst}) reveals the parameter that microscopy does not provide --- the elusive to the naked eye statistics of structural correlations, which are able to distinguish and characterize randomly looking, or {\it disordered}, and {\it permeable} samples such as those in Fig.~\ref{Fig:Samples}c, using a low-resolution bulk transport measurement. As most building blocks of living tissues, such as cells and organelles, are not fully confining (cells have permeable walls; water can move along the dendrites and axons), we believe this fundamental result can serve as a basis for  quantitative investigations of $\mu$m-level structural correlations in complex materials \cite{torquato2013random} and in live biological tissues \cite{novikov2011random,novikov2014revealing,Burcaw201518,fieremans2016} with diffusion NMR and Magnetic Resonance Imaging. 

%%%%%%%%%%%%%%%%%%%%%%%%%%%%%%%%%%%%%%%%%%%%%%%%%%%%%%%%%%%%%%%
%%%%%%%%%%%%%%%%%%%%%%%%%%%%%%%%%%%%%%%%%%%%%%%%%%%%%%%%%%%%%%%
%%%%%%%%%%%%%%%%%%%%%%%%%%%%%%%%%%%%%%%%%%%%%%%%%%%%%%%%%%%%%%%
%%%%%%%%%%%%%%%%%%%%%%%%%%%%%%%%%%%%%%%%%%%%%%%%%%%%%%%%%%%%%%%
%%%%%%%%%%%%%%%%%%%%%%%%%%%%%%%%%%%%%%%%%%%%%%%%%%%%%%%%%%%%%%%
%%%%%%%%%%%%%%%%%%%%%%%%%%%%%%%%%%%%%%%%%%%%%%%%%%%%%%%%%%%%%%%
%%%%%%%%%%%%%%%%%%%%%%%%%%%%%%%%%%%%%%%%%%%%%%%%%%%%%%%%%%%%%%%
%{\bf Acknowledgments.} 
\begin{acknowledgments}
A.P. acknowledges Steven Morgan, Farhana Gul-E-Noor and Basant Dhital for usefull discussions regarding the experimental findings. G.S.B. acknowledges support from the NIH award 2SC1GM086268. EF and DSN were supported by the Fellowship from Raymond and Beverly Sackler Laboratories for
Convergence of Physical, Engineering and Biomedical Sciences, by the Litwin Foundation for Alzheimer's Research, and by the NIH/NINDS award R01NS088040. 
\end{acknowledgments}
\bigskip

%{\bf Author contributions.}
%AP, DSN, EF and GSB designed the experiment; AP performed the experiment; GSB supervised the experiment; AP performed data analysis; AP and DSN wrote the manuscript. 
%All authors discussed the results and implications and commented on the manuscript at all stages.
%}

\bibliography{correlations}
\end{document}